\documentstyle[12pt]{article}
\input{epsf}
\newlength{\extraspace}
\newlength{\extraspaces}
\newcommand{\be}{\begin{equation}}
\newcommand{\ee}{\end{equation}}
\newcommand{\bea}{\begin{eqnarray}}
\newcommand{\nn}{\nonumber}
\newcommand{\eea}{\end{eqnarray}}

\newcommand{\nk}{\noindent}

\def\lsim{\mathrel{\rlap {\raise.5ex\hbox{$ < $}}
{\lower.5ex\hbox{$\sim$}}}}

\def\gappeq{\mathrel{\rlap {\raise.5ex\hbox{$>$}}
{\lower.5ex\hbox{$\sim$}}}}
\def\lappeq{\mathrel{\rlap{\raise.5ex\hbox{$<$}}
{\lower.5ex\hbox{$\sim$}}}}

\begin{document}

\begin{titlepage}
\begin{flushright}
CERN-TH/97-119 \\
CTP-TAMU-26/97 \\
ACT-09/97 \\
OUTP-97-28P \\
hep-th/9706125 \\
\end{flushright}
\begin{centering}
\vspace{.1in}
{\large {\bf $M$ Theory from World-Sheet Defects in Liouville
String}} \\
\vspace{.3in}
{\bf John Ellis$^{a} $},
{\bf N.E. Mavromatos$^{b,\diamond}$},
{\bf D.V. Nanopoulos}$^{c,d,e}$ \\

\vspace{.2in}
\vspace{.1in}
{\bf Abstract} \\
\vspace{.3in}
\end{centering}
{We have argued previously that black holes may be represented
in a $D$-brane approach by monopole and vortex defects in a
sine-Gordon field theory model of
Liouville dynamics on the world sheet. Supersymmetrizing
this sine-Gordon 
system, we find critical behaviour in 11 dimensions,
due to defect condensation that is the world-sheet analogue
of $D$-brane condensation around an extra space-time dimension in $M$
theory. This supersymmetric description of Liouville 
dynamics has a natural embedding within a 12-dimensional framework
suggestive of $F$ theory.}
\vspace{0.5in}
\nk $^a$ Theory Division, CERN, CH-1211, Geneva, Switzerland,  \\
$^b$
University of Oxford, Dept. of Physics
(Theoretical Physics),
1 Keble Road, Oxford OX1 3NP, United Kingdom,   \\
$^{c}$ Center for
Theoretical Physics, Dept. of Physics,
Texas A \& M University, College Station, TX 77843-4242, USA, \\
$^{d}$ Astroparticle Physics Group, Houston
Advanced Research Center (HARC), The Mitchell Campus,
Woodlands, TX 77381, USA. \\
$^{e}$ Academy of Athens, Chair of Theoretical Physics, 
Division of Natural Sciences, 28 Panepistimiou Avenue, 
Athens 10679, Greece. \\
$^{\diamond}$ P.P.A.R.C. Advanced Fellow.

\vspace{0.01in}
\begin{flushleft}
CERN-TH/97-119 \\
CTP-TAMU-26/97 \\
ACT-09/97 \\
OUTP-97-28P \\
June 1997\\
\end{flushleft}
\end{titlepage}
\newpage

\section{Introduction}

If one is to understand the relationships between the plethora
of apparently consistent classical string vacua and
understand non-perturbative effects in string theory, one needs an
approach that goes beyond critical strings and conformal field
theory. Important ingredients in such an understanding are
provided by string soliton configurations such as $D$ 
branes~\cite{dbranes}.
These are powerful tools that have facilitated the deeper understanding of
string theory that is now provided by duality, which links Type I, IIA,
IIB and heterotic strings within a unified framework that
takes its name from the 11-dimensional limit called $M$ theory~\cite{duff}.
Some string dualities may find a clearer interpretation within
a possible 12-dimensional extension called $F$ theory~\cite{ftheory}.

Another approach introduces a Liouville field on the world sheet,
which enables one to construct critical string theories in unusual
numbers of dimensions~\cite{gervais,aben,ddk}. It was 
also shown that this approach could
be used to introduce classical string vacua with
interpretations as non-trivial space-time backgrounds such as
Friedmann-Robertson-Walker cosmologies~\cite{aben}. This approach was
subsequently extended to the quantum level~\cite{emn}, where it was argued
that higher-genus and non-perturbative effects would deform the
theory away from classical criticality, inducing non-trivial
dynamics for the Liouville field.

This development was motivated by, and used to accommodate,
processes involving string black holes in $1+1$ dimensions~\cite{emn}. In
particular, the creation and disappearance of such quantum black holes,
as well as transitions between them that correspond to back-reaction of
particles on the black-hole background metric, could be encoded in the
dynamics of the Liouville field. A representation for this dynamics
was found~\cite{emn} in 
terms of defects on the world sheet: vortices and monopoles
of the Liouville field that could be described in terms of a deformed
sine-Gordon model~\cite{ovrut}.

More recently, it has been shown that $D$ branes emerge naturally
from Liouville string theory~\cite{emnd,diffusion}. 
Quantum effects at higher genus
introduce fluctuations in the 
$\sigma$-model couplings of the conformal field
theory that describes a classical string vacuum. These are  associated
with singular configurations in the space of higher-genus moduli, such
as long, thin handles that resemble wormholes on the world sheet.
The absorption of the divergences associated with these singularities
requires quantization of the world-sheet couplings and 
the introduction of $\sigma$-model
counterterms that have a space-time interpretation as $D$ 
branes~\cite{emnd,diffusion}.
Moreover, $D$ branes in {\it arbitrary} space-time
dimensions can be described  within this approach 
in terms of vortex and monopole defects
in the Liouville theory on the world sheet~\cite{emnmonopdbr}, 
just as was the case for
$(1+1)$-dimensional black holes. 
When these defects are irrelevant operators in a
renormalization-group sense, the world sheet is in a dipole phase,
and their correlators
with the vertex operators of target-space matter excitations have
cuts, in general. Thus the world sheet becomes
effectively open, and $D$ branes 
arise when suitable boundary conditions are
implemented along the cuts~\cite{emnmonopdbr}.
This provides a $D$-brane description of quantum fluctuations
in the space-time background, i.e., space-time foam, with 
microscopic black holes represented by $D$ branes whose
appearance and disappearance is described by defect
fluctuations on the world sheet.

The next step in this programme, described in this paper, is to
supersymmetrize the world-sheet sine-Gordon theory that describes
the non-critical Liouville dynamics~\cite{ovrut,superliouville}. 
We enumerate the phases of
sine-Gordon models with $n=1,2$ world-sheet supersymmetry,
identifying the ranges of space dimensions where their world-sheet
vortices and monopoles can be classified as relevant or irrelevant
deformations, focussing on the vortices and monopoles of lowest
charge, which are constrained by quantization conditions. In the
case of $n=1$ world-sheet supersymmetry, we show that the lowest-charge
$|q| = 1$ vortices are irrelevant below $d=11$
dimensions, as seen in Fig.~1. Likewise,
the lowest-charge $|e| = 1$ monopoles are irrelevant above
$d=17$ dimensions. Normal space time with massive black holes is
described by the phase in which vortices are irrelevant.
There are Kosterlitz-Thouless~\cite{KT}
(KT) phase transitions in $d = 11, 17$ dimensions, with the
new phenomenon of defect condensation and an unstable
plasma phase for intermediate dimensions.
The appearance
of massless defect configurations in $d=11, 17$ marks recovery of
conformal symmetry. Away from these critical condensation points,
one has effectively a $D = (d +1)$-dimensional target theory, but 
the extra dimension decouples in the conformal theory when $d=11$, just
like the Liouville field in conventional critical string theory.

We interpret the transition of world-sheet vortices from
irrelevancy to relevancy in 11 dimensions as the world-sheet
transcription of the masslessness of $D$ branes in the
strong-coupling $11$-dimensional limit of $M$ theory. 
We also discuss how the theories with $n=1$ world-sheet
supersymmetry that we discuss may be described as marginal
deformations of theories with $n=2$ world-sheet supersymmetry~\cite{baulieu},
thereby making more direct contact with string models with
$N=1$ space-time supersymmetry. Finally, we
conjecture that the 
extra dimension provided by the Liouville
field with non-trivial dynamics 
may be interpreted as a world-sheet description of $F$
theory~\cite{ftheory}.

\section{Sine-Gordon Description of World-Sheet Vortices,
Monopoles and $D$ Branes}

The world-sheet sine-Gordon action~\cite{kogsath,ovrut} of such defects 
in a purely bosonic theory was used 
in~\cite{emn} to describe $(1+1)$-dimensional black
holes, and in~\cite{emnmonopdbr}
to describe target-space $D$ branes. For completeness and
to set the scene for the supersymmetric extension we use later,
we briefly review here the bosonic 
sine-Gordon formalism of~\cite{ovrut}, whose
$\sigma$-model partition function is:
\bea
Z&=&\int D{\tilde X} exp(-\beta S_{eff}({\tilde X}) )  \nn \\
\nonumber
\beta S_{eff}&=&  \int d^2 z [ 2\partial {\tilde X}
{\overline \partial } {\tilde X} +  
\frac{1}{4\pi }
[ \gamma _v\omega ^{\frac{\alpha}{2}-2}
(2 \sqrt{|g(z)|})^{1-\frac{\alpha}{4}}: cos (2\pi\sqrt{\beta}q 
[{\tilde X}(z) + {\tilde X}({\bar z})]):   \\
& +&   
\frac{1}{4\pi }
[ \gamma _v\omega ^{\frac{\alpha'}{2}-2}
(2 \sqrt{|g(z)|})^{1-\frac{\alpha'}{4}}: cos (\frac{e}{\sqrt{\beta}}
[{\tilde X}(z) - {\tilde X}({\bar z})]):] ]
\label{sevenv}
\eea
where 
$\alpha \equiv 2\pi\beta q^2$, $\alpha' \equiv e^2/2\pi\beta$, with
$q, e$ the charges of the vortex and monopole solutions:
\be
\partial _z\partial _{\bar z} {\tilde X}_v =i\pi \frac{q}{2}
[\delta (z-z_1)-\delta(z-z_2)],  \qquad 
   \partial _z  \partial _{\bar z} {\tilde X}_m =-\frac{e \pi} {2}
[\delta (z-z_1) -\delta (z-z_2)]
\label{sourcev}
\ee
Moreover, 
$\omega$ is an ultraviolet angular cutoff in (\ref{sevenv}), 
and $\gamma_{v,m}$ are the 
fugacities for vortices and monopoles.
We note that $\beta^{-1}$ plays the r\^ole of an effective 
temperature in (\ref{sevenv}), 
and that the vortex and monopole 
operators have anomalous dimensions: 
\be
      \Delta _v =\frac{\alpha }{4}=
      \frac{\pi\beta}{2}q^2, 
\qquad 
   \Delta _m =\frac{\alpha '}{4}=
      \frac{e^2}{8\pi\beta}
\label{confdim}
\ee 
The system (\ref{sevenv}) 
is invariant under the $T$-duality transformation~\cite{ovrut}: 
$\pi \beta \leftarrow\rightarrow \frac{1}{4\pi\beta}~;~q 
\leftarrow\rightarrow e$. 

We identify ${\tilde X}$ in (\ref{sevenv}) 
with the rescaled Liouville field $\sqrt{\frac{C-25}{3g_s^\chi}}\phi$ -
where $g_s$ 
denotes the string coupling, and $\chi$ is the Euler characteristic
of the world-sheet manifold - after subtraction of
classical solutions to the equations of motion and  
spin-wave fluctuations~\cite{ovrut}.
As the analysis of~\cite{emnmonopdbr}
indicates, one has $e \propto \frac{1}{\sqrt{g_s}}$. 
This is used in our analysis below, when we discuss 
the spectrum of stable configurations of the theory. 
In our approach, we relate $\beta$ to the 
Zamolodchikov $C$ function of the accompanying matter~\cite{zam}: 
$\beta \rightarrow \frac{3g_s^\chi}{\pi(C-25)}$. 

In ref. \cite{emnmonopdbr}
we were interested in the case $C > 25$, and 
we considered only irrelevant deformations,
that do not 
drive the theory to a 
new fixed point. The world-sheet  
system was then found to be in a dipole phase~\cite{ovrut},
and had the general features discussed in the context of 
the formulation of quantum non-critical Liouville
string in~\cite{emn}, namely non-conformal 
deformations,
non-trivial Liouville dynamics, and dependence 
on an ultraviolet cutoff. We interpreted this system 
in~\cite{emn} as a model 
for space-time foam in the context of $(1+1)$-dimensional
string black holes. 

We further suggested in~\cite{emnmonopdbr} that the
field theory (\ref{sevenv}, \ref{sourcev}) also represents
$D$-brane foam.
This is because the correlation functions 
of these non-conformal deformations
with other (conformal) deformations,
representing vertex operators for the excitation
of matter fields in target space, 
appear to have cuts, thereby leading to effectively open world sheets.
Consider, for example, the scattering of closed string states
$V_T(X)=exp(ik_M X^M - iEX^0): 
M=1, \dots D_{cr}-1$, where $D_{cr}$ is the 
critical space-time dimension, in the presence 
of a monopole defect. An essential aspect of this
problem is the singular behaviour of the operator product
expansion of $V_T$ and a vortex or monopole operator $V_{v,m}$. Treating
the latter as a sine-Gordon deformation of (\ref{sevenv}),
computing at the tree level using the free world-sheet
action, and suppressing for brevity anti-holomorphic parts,
we find
\bea
&~&\hbox{Lim}_{z \rightarrow w}<V_T (X^0,X^i)(z) 
V_{v,m} (X^0)(w) \dots > \nn \\
&~&\sim \int d^{D_{cr}-1}k \int dE \delta^{D_{cr}-1} (k) 
(\dots) 
[\delta (\Sigma E + \kappa/\sqrt{\beta} ) (z-w)^{-\Delta _T - \Delta _{v,m}} + 
\nn \\
&~&\delta (\Sigma E - \kappa/\sqrt{\beta} ) (z-w)^{-\Delta _T - \Delta _{v,m}} ]  
\label{twopoint}
\eea
where $\kappa = 2\pi\beta q$ for vortices,  $\kappa = e$
for monopoles,  
$\Delta _T =\frac{E^2}{2}$, 
the energy-conservation $\delta$ functions result from integration
over the Liouville field ${\tilde X} \equiv X^0$, and
$(\dots)$ indicates factors 
related to the spatial momentum components of 
$V_T$, other vertex operators in the correlation function, etc..  
We see that (\ref{twopoint}) 
has cuts for generic values of 
$\Delta_T + \Delta _{v,m}$, causing the theory 
to become effectively that of an open string. 

To relate this observation more closely to $D$ branes,
we recall that there are two possible types of
conformal boundary conditions, Dirichlet and Neumann,
related to each other by appropriate 
$T$-duality transformations, which in the above picture
appear as canonical transformations in the path 
integral~\cite{otto,emnmonopdbr}. Thus
one may place Dirichlet boundary conditions 
on the boundaries of the effective world sheet, i.e., along the 
cuts, thereby obtaining solitonic 
$D$-brane~\cite{dbranes} configurations.
Some non-trivial consistency checks of this identification
of world-sheet defects with target-space $D$ branes,
including derivations of the correct energy and momentum conservation
relations during the scattering 
of light closed-string states off these solitons, have been 
obtained in~\cite{emnmonopdbr}, using the identification
of the zero mode of the Liouville field with the target time.
We interpret the resulting world-sheet vortex/space-time $D$-brane
system as a representation of space-time foam and microscopic
black holes in higher dimensions, as a natural extension of the
previous (1+1)-dimensional approach.

\section{Critical Dimensions in Supersymmetric Sine-Gordon
World-Sheet Theory}

In this section we follow the discussion of~\cite{ovrut}
on the supersymmetrization of the monopole and vortex 
configurations on the world sheet of the string, with a 
view to 
relating their physical interpretation to $M$ theory.
We start with a sine-Gordon theory with local $n=1$ world-sheet 
supersymmetry, whose (non-chiral) monopole deformation is~\cite{ovrut}:
\be
     V_m = {\overline \psi}\psi 
:{\rm cos}[\frac{e}{\beta_{n=1}^{1/2}}(\phi(z)- \phi ({\overline z}))]:
\label{monopole}
\ee
where the
${\overline \psi}, \psi$ are world-sheet fermions, which transform as 
fields of conformal dimensions $(0,\frac{1}{2}), (\frac{1}{2},0)$,
respectively, $\phi $ is the Liouville field, which 
corresponds as in~\cite{ovrut} and above to the
fluctuating quantum part of the 
conformal scale factor of the original world-sheet 
metric with the `spin-wave' 
and classical parts appropriately subtracted, and
$:~~:$ denotes normal ordering.
The equivalence of the above $\sigma$ model with 
Liouville theory implies that 
the effective `temperature' $1 / \beta_{n=1}$ is related 
to the matter central charge $d$~\cite{ovrut} by:
\be
\beta_{n=1}=\frac{2}{\pi (d-9)}
\label{efftemprn1}
\ee
where we assume the case $d > 9$. 
The `matter' central charge $d$ may be considered as the 
number of dimensions  of an effective 
target space-time in which the theory lives. 
The deformation (\ref{monopole}) is understood to be added 
to the free-fermion and Liouville actions, and the combined system 
has $n=1$ world-sheet supersymmetry. 
The corresponding deformation for vortex solutions is given by:
\be
     V_v = {\overline \psi}\psi 
:{\rm cos}[2\pi q\beta_{n=1}^{1/2}(\phi(z)+ \phi ({\overline z}))]:
\label{vortex}
\ee
where $q$ is the vortex charge. For simplicity, we 
consider the case where the world-sheet cosmological constant 
is zero.

Including the conformal dimensions $(\frac{1}{2},0), (0,\frac{1}{2})$
of the fermion fields, 
the conformal dimension $\Delta_v$, $\Delta_m$ 
of the vortex and monopole deformations (\ref{vortex}),(\ref{monopole}) 
in the holomorphic sector are,
respectively:
\bea
&~&  \Delta_v=\frac{1}{2} + \frac{1}{2}\pi \beta_{n=1} q^2 = 1/2 +
q^2/(d-9) \nn \\
&~& \Delta_m=\frac{1}{2} + \frac{1}{8\pi \beta_{n=1}}e^2=1/2 + e^2 (d-9)/16 
\label{anomdim}
\eea
As usual, relevant deformations have $\Delta_{m,v} < 1$ and irrelevant 
deformations $\Delta _{m,v} > 1$, whilst marginal deformations have
$\Delta _{m,v} =1$. In the relevant case, the vacuum
is unstable with respect to condensation of the corresponding
topological world-sheet defects, and the system dissociates into a 
plasma of the corresponding free charges, whilst in the irrelevant case
the vacuum  is stable with respect to condensation of the corresponding
defects. Marginal deformations correspond to a Kosterlitz-Thouless
phase transition~\cite{ovrut,KT}, and respect the conformal 
invariance of the system.

An important constraint on the charges of the monopole and vortex 
configurations is imposed by the {\it single-valuedness}
of the partition function on a spherical world
sheet~\cite{ovrut}. We recall
that the monopole and vortex configurations are related by a 
$T$-duality transformation which corresponds to a temperature 
inversion~\cite{ovrut}:
\be 
        e \leftarrow\rightarrow q \quad; \quad \pi\beta \leftarrow\rightarrow 
\frac{1}{4 \pi \beta} 
\label{duality}
\ee
This implies the co-existence of 
{\it both} deformations if either one is present.
Single-valuedness of the partition
function then requires the following relation between the
monopole and vortex charges:
\be
   2\pi \beta_{n=1}q = 4q/(d-9)= e \qquad q,e \in Z
\label{quantization}
\ee
It is easy to verify that this condition imposes equality
between the conformal dimensions (\ref{anomdim}) of the vortex
and monopole defects: $\Delta_v = \Delta_m$.

\begin{figure}[htb]
\epsfxsize=5in
\centerline{\epsffile{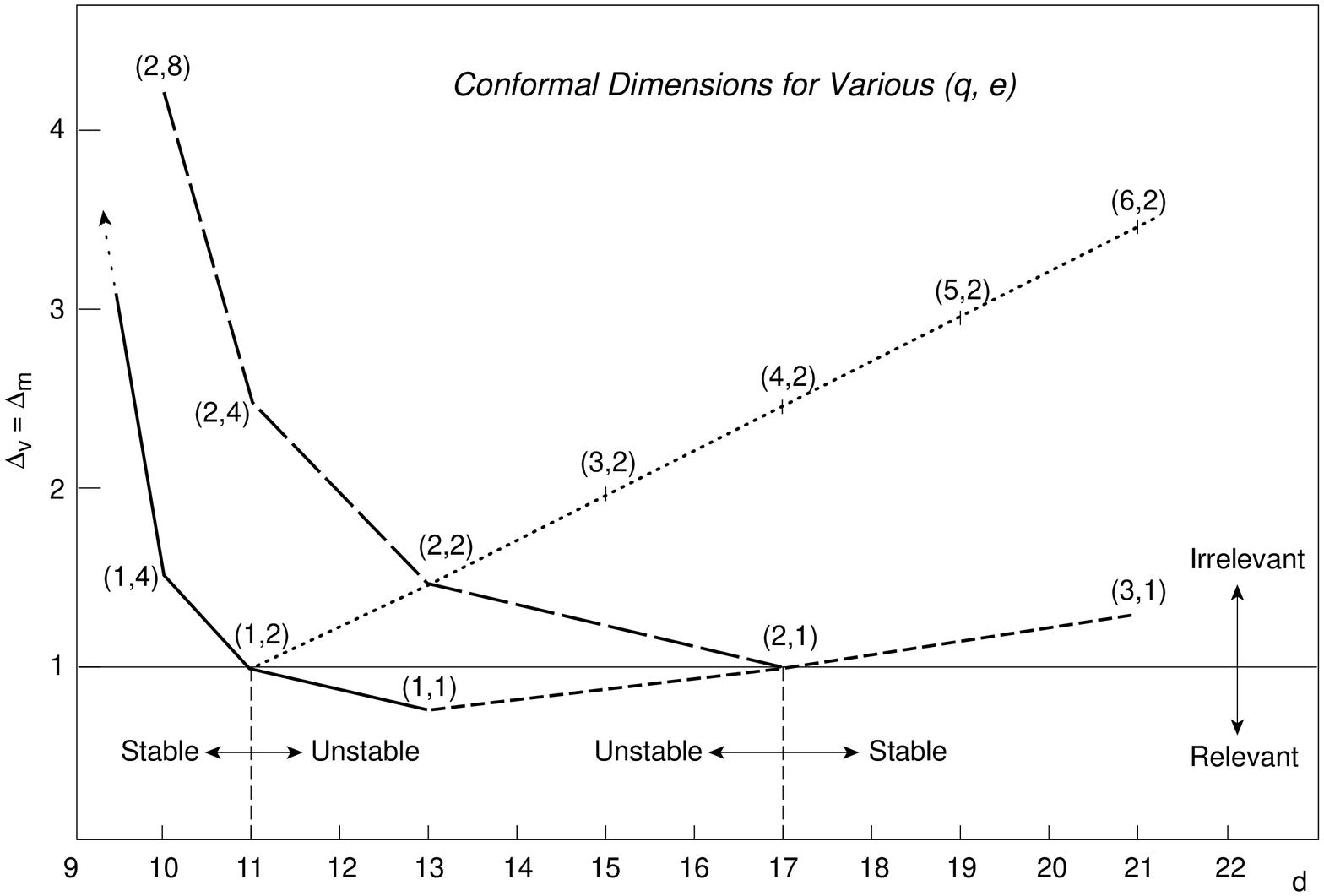}}
\caption{{\it Chart of the conformal dimensions of $n=1$ world-sheet
supersymmetric sine-Gordon defects in integer dimensions $9 < d < 22$:
the operators are marginal for $\Delta_v = \Delta_m =1$.
Vortices with $|q| = 1$ are linked by solid lines, vortices with $|q| = 2$ by
long-dashed lines, monopoles with $|e| = 1$ by short-dashed lines, and
monopoles with $|e| = 2$ by dotted lines. In each case, the charges of
the $T$-dual defects given by 
the quantization condition (\ref{quantization}) are also indicated.
Note that the world-sheet vacuum is stable for $d < 11$ and $d > 17$,
and dissolves into an unstable plasma phase of free $|q| = 1$
and/or $|e| = 1$ defects at intermediate dimensions.}}
\end{figure}

We consider first the 
case with the lowest possible vortex charge $|q|=1$. We
see from (\ref{anomdim}) that defect deformations are
irrelevant, and the vacuum is stable, if 
\be
        0 <  d-9 < 2q^2 =2 \rightarrow  9 < d < 11 
\label{conditionvortex} 
\ee
The quantization condition (\ref{quantization}) then
implies that the allowed monopole charges
in this case are:
\be 
  e=4q/(d-9)=\pm 4/(d-9)
\label{conditionmono}
\ee
The vacuum for this monopole is {\it also} stable in the range
(\ref{conditionvortex}), since this
monopole deformation (\ref{monopole}) is also irrelevant,
with the same conformal dimension as the lowest-lying vortex.
The conformal dimensions of vortices with $|q| = 1$ (and of
the corresponding monopole charges $e$) are shown in Fig.~1
for allowed integer values of $d > 9$, linked by solid lines
to guide the eye.

Some comments are in order about defects with larger charges.
If one restricts oneself to the 
region $9 < d < 11$~\footnote{We recall that $d$ is the central charge of
the matter system, which need not be integer.},
higher-charge monopole configurations with 
\be 
        e^2 > \frac{8}{d-9} \qquad e \in Z
\label{irel}
\ee
are irrelevant deformations.
In the particular case of standard critical strings with $d=10$,
the stable monopole configurations are those with charges 
$e > 2\sqrt{2}$, i.e., the integers $|e|=3,4,5 \dots$, whilst for $d=9$ 
all finite $|e|$ configurations are unstable. 
As already mentioned, we see from
the analysis of~\cite{emnmonopdbr} that one can identify 
$1 / e^2 \propto g_s $, the string coupling 
in the case of a $D$-brane $\sigma$-model. Thus the above 
infinite tower of stable discrete monopole configurations
corresponds to an infinite
tower of massive $D$-brane states, 
with masses $\propto e^2$ in appropriate units.
The conformal dimensions of $|q| = 2$ vortices (and the
corresponding monopoles) are shown in Fig.~1, linked by long-dashed
lines to guide the eye, and we see explicitly that they are
irrelevant for all $d$.
  
Consider now the case of the lowest monopole charge
$e=\pm 1$, when
the quantization condition (\ref{quantization}) 
implies that the allowed vortex charges are:
\be
      q=\pm (d-9)/4
\label{conditionvort}
\ee
The corresponding defect operators are both irrelevant, and the
vacuum stable, for 
\be 
         d > 17 
\label{conditionmonopol}
\ee
The conformal dimensions of $|e| = 1$ monopoles are
shown in Fig.~1, linked by short-dashed lines to guide the eye.
Also shown in Fig.~1 are the conformal dimensions of $|e| = 2$ monopoles,
which are never relevant, linked by dotted lines.

The monopole and vortex operators are {\it both} 
marginal, and the theory conformal, for: 
\be 
     {(d-9)e^2 \over 16} = {q^2 \over (d-9)} = {1 \over 2}
\label{marginal}
\ee
which, on account of the quantization condition (\ref{quantization}),
singles out the two limiting cases:
\bea 
d=11, |q|=1, |e|=2; \,\,\, d=17, |e|=1, |q|= 2
\label{marginalunique}
\eea
which are clearly visible in Fig.~1.
At these points, the monopole and vortex configurations are 
both at their 
Kosterlitz-Thouless (KT) transition points~\cite{KT}. 
For $11 < d < 17$, the world sheet is in an unstable plasma phase
with free defects.

The world-sheet cuts generated by the non-integer
anomalous dimensions in (\ref{twopoint}), which lead
to effectively open world sheets in the construction 
of~\cite{emnmonopdbr}, disappear at the conformal points
(\ref{marginalunique}), and the above
world-sheet analysis is no longer valid. 
Instead, at a KT point, the system of defect dipoles 
dissociates into a plasma of free charges, and 
the world-sheet system undergoes a phase transition. 
At these KT points the infinite tower of discrete massive $D$-brane states
(\ref{irel}) disappears~\footnote{This transition bears some analogy with
the finite-temperature deconfinement transition in QCD, where an infinite
set of massive bound states disappears.}. 

\section{Physical Interpretation of Vortex Condensation as a 
Representation of $M$~Theory.}

We now seek to relate this discussion to the manner in which
the eleventh dimension first appeared in $M$ theory, namely
via $D$-brane condensation in the strong-coupling
limit of conventional 10-dimensional strings with space-time
supersymmetry~\cite{dbranes,duff}. We recall that, like any other
solitons, the
target-space masses of $D$ branes $m_D \sim 1/g_s$, where $g_s$ is
the string coupling. These masses are reminiscent of those of
conventional string winding modes if the winding radius
$R \rightarrow \infty$ as $g_s \rightarrow \infty$. Thus
$D$-brane condensation as $m_D \rightarrow 0$ was shown to
correspond to the appearance of an eleventh large dimension,
leading to a theory whose low-energy field-theoretical limit
was 11-dimensional supergravity.

We have shown previously how $D$ branes appear naturally
in the world-sheet Liouville $\sigma$-model approach to non-critical
string theory~\cite{emnd}. Moreover, we have presented arguments that
vortex defects in the world-sheet Liouville field theory provide
an explicit representation of $D$ branes~\cite{emnmonopdbr}. Furthermore,
we have identified in the previous section two critical dimensions $d^*$
of the $n=1$ supersymmetric sine-Gordon model, where
the world-sheet defects interpreted as target-space
black holes and $D$ branes condense, namely $|q| = 1$ vortices in $d^* =
11$ and $|e| = 1$ monopoles in $d^* = 17$. We claim that this provides
an explicit world-sheet representation of $D$-brane
condensation.

It is natural to seek to identify the 11-dimensional case with 
the corresponding critical limit of $M$
theory: quite apart from anything else, the 17-dimensional case
has no known field-theoretical limit. However, there are two 
apparent puzzles
to be resolved before this putative interpretation 
of the $d^* = 11$ theory can be established. 

One apparent puzzle is that in non-critical string theory we identify
the Liouville field with an extra target space-time dimension, 
which would seem to provide us with $d^* + 1 = 12$ dimensions.
This `non-critical' target space time must have indefinite signature,
with at least one time-like dimension provided by 
the Liouville field construction~\cite{aben},
and the fact that $d > 9$. However, we recall that
both the $|q| = 1$ vortex and $|e| = 2$ monopole
deformations are exactly marginal at $d^* = 11$, so
that the $n = 1$ supersymmetric version of (\ref{sevenv})
is conformal. In such a case, as in the case of
conventional critical string theory, the
background target-space time fields $g^i$,
that deform the $\sigma$-model when coupled to the 
appropriate vertex operators, depend only on 
the $d^*$ coordinates, due to the conformal nature of the pertinent 
$\sigma$ model. One way to see this is to 
recall that the Liouville 
mode acts in general as a covariant local renormalization scale 
on the world sheet of the non-critical string~\cite{emn},
but that this scale is redundant for a critical theory.
Thus the effective target-space field theory at $d^*$ 
is completely characterized by fields that are
independent of the Liouville field. Hence the supersymmetric
sine-Gordon theory at the KT critical point is effectively
an 11-dimensional theory, and may be identified with $M$ theory.

Of course, the simple sine-Gordon model (\ref{sevenv}) and its
$n = 1$ supersymmetric extension do not manifest all the mathematical
structure of string theory, and hence $M$ theory. This should be
reflected in consistency conditions on the possible choices of matter
and space-time degrees of freedom that must be satisfied by any
critical string construction. Our identification of vortex
condensation in $n = 1$ sine-Gordon Liouville theory should be
understood by analogy with the $Z_3$ Potts model description of the
finite-temperature QCD deconfinement phase transition. The reduced system
captures the essence of the critical behaviour of the full system.

The second apparent puzzle is that, since we have been working with
$n = 1$ supersymmetric sine-Gordon theory on the world sheet, we do not
have $N = 1$ supersymmetry in target space, which would require $n = 2$
supersymmetry on the world sheet. There is 
also the issue of the signature 
of the $d^*$ dimensions, which
can be resolved in a supersymmetric target-space theory.
The next section is devoted to a discussion of these two issues.

\section{Relation to Models with $N = 1$ Space-Time Supersymmetry}

As already mentioned, local $n=1$ world-sheet supersymmetry does
not admit a supersymmetric target
space-time interpretation, for which one needs $n=2$ 
local world-sheet supersymmetry. 
In the context of the monopole and vortex deformations 
discussed in earlier sections, an $n=2$ 
locally supersymmetric sine-Gordon model may easily 
be constructed~\cite{ovrut},
as a straightforward generalization of the $n=2$ super-Liouville 
theory~\cite{superliouville}. 
The latter originates from a world-sheet 
$n=2$ supergravity (local supersymmetry). The simplest 
model~\cite{superliouville} is 
the  $O(2)$-symmetric $n=2$ supergravity, which is a
two-dimensional theory admitting the string
interpretation~\cite{ademollo}, in the sense 
of giving rise to a consistent four-dimensional target space-time.
The particle content of the $n=2$ supergravity multiplet is
one graviton (zweibein) field $e^a_\mu$, 
two Majorana or one Dirac gravitino 
$\chi^\mu$, and a vector field $A_\mu$. The supergravity multiplet
is coupled to a set of $D$ complex matter multiplets, consisting of a complex
scalar $X$, the Dirac spinor $\psi$ and a complex auxiliary field $F$. 
We do not give here details of the action, since
we only use the theory for illustrative purposes:
details may be found in~\cite{n=2sugra}. 

The Liouville dressing of this theory is facilitated by picking
the conformal gauge: $e^a_\mu=e^\phi {\hat e}^a_\mu$, 
$\chi_\mu=\frac{1}{2}\gamma_\mu(\eta_1 + i\eta_2)$, 
$A_\mu=\frac{1}{2}\epsilon_{\mu\nu}\partial^\nu \rho$. One has also 
gauge-fixing and reparametrization Fadeev-Popov ghosts, which contribute
$c_{gh}=-6$ to the central charge.
The ghost sector of the $n=(2,2)$  world-sheet local supersymmetric 
string consists of the $(b,c)$ ghosts for general coordinate
invariance, contributing (-26, -26) to the central charge, 
the two bosonic ghosts of $n=(2,2)$ supersymmetry:
$(\beta^1,\gamma^1), (\beta^2,\gamma^2)$, each of conformal weight
(3/2,-1/2) and contributing a total of (22,22) to the central charge,
and the fermionic ghost fields $(f,g)$ of the 
local $O(2) \simeq U(1)$ gauge algebra of the $N=2$ superconformal
symmetry, 
which have conformal weight (1,0) and contribute (2,2) to the conformal 
anomaly. The total contribution of the ghost sector to the 
central charge is, therefore, $c_{gh}=-6$, implying that 
under conformal transformations the effective ghost-induced 
gravity action transforms as:
$S_{gh} \rightarrow S_{gh} + \frac{1}{48\pi}c_{gh}S_{sl}$,
where $S_{sl}$ is the super-Liouville $n=2$ $O(2)$ 
action~\cite{superliouville}:
\be
S_{sl}=\frac{1}{2\pi}\int d^2z ({\overline \partial}\phi 
\partial \phi - {\overline \partial}\rho \partial \rho 
+ \sqrt{g} R^{(2)} \phi -  \sqrt{g} R^{(2)} \rho -  
\sum_{i=1,2} [\eta_i {\overline \eta_i} - {\overline \eta}_i \partial 
{\overline \eta}_i ] )
\label{slaction}
\ee
Notice the wrong sign of the second scalar field $\rho$, which indicates
that this field is a ghost. Going to a fiducial metric, one can 
cast the above action in a more  familiar  form where the 
central charge deficit $Q=\sqrt{\frac{1-d}{2}}$, 
with $d$ the `matter' central charge,  
appears in front of the 
curvature terms as usual. The super-Liouville theory can be made physical,
by appropriate Wick rotations of the $\phi, \rho$ fields, for 
{\it any } $d$, even $d > 1$.  This is due to the fact that the fields 
$\rho$ and $\phi$, since their kinetic terms have opposite signs,
simply interchange their r\^oles as one crosses the 
value $d=1$~\cite{superliouville}. 

The presence of monopoles does not alter the above features. 
In particular, the monopole deformation of the $n=2$ theory, which 
respects $n=2$ local supersymmetry of the conformally-fixed action,
reads~\cite{ovrut}:
\be
      {\overline \eta}_1 \eta_1 {\overline \eta}_2 \eta_2 
:cos[\frac{e}{\beta_{n=2}^{1/2}}(\phi(z)-\phi ({\overline z}))]:
\label{n=2monopole}
\ee
and there is a corresponding expression for the vortex, with
\be
   \beta_{n=2}=\frac{4}{\pi (1-d)}
\label{betan=2}
\ee
The fields $\eta_i$ in (\ref{n=2monopole}) are conformal fields of
dimension (1/2, 0),
so that the above $n=2$ monopole deformation has conformal dimension 
\be
     \Delta_{n=2,m} \ge 1  
\label{n=2conf}
\ee
in each sector. Therefore,
the monopole and vortex deformations are {\it irrelevant} 
for every $d$, in accord with the above statement that the 
super-Liouville theory is defined for every $d$. 
The value $d=2$ corresponds to the critical four-dimensional target
space-time string of~\cite{ademollo}, for which 
the Liouville fields decouple. 

To understand the relation of $n=2$ world-sheet supersymmetric 
models to $n=1$, we note that 
this $n=2$ string has been shown recently to be 
connected by world-sheet {\it marginal} deformations to 
strings with less world-sheet local supersymmetry, such as (1,1) or (1,0) 
superstrings~\cite{baulieu}. This
was achieved by adding suitable topological 
packages to the critical fields of the $(2,2)$ $n=2$ world-sheet theory,
whose action was described by a $BRST$-exact form.
In the case of (2,2) string, 
such packages 
consist of fermionic quartets $(\lambda_i^\alpha, \rho_i^\alpha)$,
$\alpha=1,\dots 4$, and their bosonic partner fields 
$(F_i^\alpha, {\overline F}_i^\alpha)$, where $i$ runs over
appropriate packages.
These fields have conformal weights (1,0), and
their world-sheet $BRST$-exact action is 
\be
   I_{top}=i\int d^2z \sum_{\alpha=1}^4 Q_{BRST}({\overline F}_i^\alpha 
{\overline \partial} \rho_i^\alpha)
\label{topaction}
\ee
with the $BRST$ transformation $Q_{BRST}\rho_i^\alpha =iF_i^\alpha$,
$Q_{BRST}F_i^\alpha=0, Q_{BRST}{\overline F}_i^\alpha=\lambda_i^\alpha$,
and $Q_{BRST}\lambda_i^\alpha=0$. Such packages do 
not introduce any anomalies, so any number of them can be added 
to the world-sheet action of the critical fields. 

World-sheet supersymmetry breaking was achieved using
marginal deformations by twisting
the gravitino ghosts $(\beta^2, \gamma^2)$ 
in such a way that their conformal weight (-3/2, -1/2)
was twisted to (1/2, 1/2), so that they should be interpreted as `matter'
fields in an $n=1$ locally-supersymmetric world-sheet theory.
The twisting is achieved by exactly-marginal deformations
of Liouville type, which deform the theory in an appropriate
way, so that the new conformal dimensions of the gravitino ghosts,
with respect to the stress tensor deformed by the marginal 
deformations, 
are the ones appropriate for the $n=1$ world-sheet 
supersymmetry~\cite{baulieu}.
The marginal deformations of the critical $d=2$, $n=2$ theory
consist of linear-dilaton terms~\cite{aben} of the
form~\cite{emnd,baulieu}
\be
     \sum_{i=1}^N \int d^2z \frac{1}{2\pi}\sqrt{g}({\overline \partial}\Phi_i 
\partial \Phi_i - i\frac{1}{2}k^i R^{(2)}\Phi_i) 
\label{lineardilaton}
\ee
where g is a world-sheet metric. 
The coefficients $k_i$ are chosen to be null:
$\sum_i k_i^2 =0$, so that the total central charge of such terms
is equivalent to that of the free bosonic system, which decouples from the 
rest of the theory~\cite{baulieu}. 
In this way the total central charge of the theory is unaffected 
by the presence of such terms. 
Compensating twists are necessary in the `topological package' sector,
for consistency. For instance, the breaking of $n=2$ to $n=1$
world-sheet supersymmetry requires~\cite{baulieu} 
that the conformal dimension (1,0)
of the $(\lambda_1^\alpha,\rho_1^\alpha)$ fields be twisted to (1/2, 1/2).
Further twists of the remaining ghost fields, and/or the ghost fields
of additional topological packages, 
can lead to further reduction of the world-sheet
supersymmetries, thereby connecting the $n=2$ string to 
other $\sigma$ models~\cite{baulieu}. 

These observations are important for our approach, since they indicate the
possibility that, even in the non-critical 
sine-Gordon Liouville case discussed 
in previous sections,
one can find marginal deformations that connect them to
theories with $n=2$ local world-sheet supersymmetry, that admit 
a supersymmetric target-space interpetation.
Supersymmetry can be broken partially by marginal 
deformations of similar linear-dilaton nature to the 
ones discussed just above in the
critical $(2,2)$ case. The ghost sector of the theory remains the same:
the only difference from the critical model of~\cite{ademollo}, for which 
the Liouville field decouples, is the fact that now the 
matter theory is more complicated. It contains in general
curved $d$-dimensional target-space background coordinates and their 
world-sheet fermionic partners, which should be $N=1$ supersymmetric in
target space. We do not enter here into the 
formal construction of such a world-sheet supergravity theory,
which does not appear necessary for our present purpose.

Since the ghost sector retains the features of~\cite{ademollo},
we observe that
the deformed stress tensor of the theory, with
respect to which the new conformal dimensions of the various fields 
are defined, can be arranged such that the gravitino ghost  
$(\beta^2, \gamma^2)$ is twisted in the same way as previously.
This will connect the $n=2$ world-sheet supersymmetric background
to the $n=1$ theory. As we have already seen, when one admits 
topologically non-trivial structures on the world sheet,
the only conformal
points of the $n=1$ super sine-Gordon model are those corresponding 
to target dimensions $d=11$ or 17. Thus we relate
strings with $N = 1$ target-space supersymmetry to
the $n = 1$ world-sheet supersymmetric representation
of the critical limit of $M$~theory discussed in previous sections.

Another way in which one can relate this $n=1$ sine-Gordon model
for world-sheet defects to the $n=2$ sine-Gordon model 
is via  condensation of the supersymmetric fermions of 
the $n=2$ super Liouville sector. To see how this may arise,
consider the Thirring-type
monopole deformation (\ref{n=2monopole}) of the $n=2$ theory 
as an example. World-sheet fermion 
interactions may lead to condensation of the $\eta_2$ fermions, 
$<{\overline \eta_2}\eta_2> =\sigma \ne 0$, which 
makes the vertex operator look like 
the $n=1$ monopole deformation (\ref{monopole}), up to
a difference $\beta_{n=2} \ne \beta_{n=1}$. 
Since we are interested only in the conformal 
points, this difference can be absorbed 
in a constant rescaling of the Liouville field in the $n=2$ theory.
The relative normalization, then, between the Liouville kinetic term 
and the monopole deformation 
can be cast into the same form as 
that of  the standard $n=1$ theory, discussed in section 3, 
by giving an appropriate value to  
the fermion condensate $\sigma$. This {\it fixed} fermion condensate
breaks the second supersymmetry, which corresponds to a $U(1)$ gauge
symmetry. 
This can be achieved in a conformal invariant way only if the 
central matter charge of the $n=2$ theory is $d=11$ or $17$.
The above procedure, therefore, describes an alternative mechanism
for relating $n=2$ and $n=1$ sine-Gordon theories without violating 
the conformal invariance of the underlying $\sigma$-model theory. 

\section{Conjectures Concerning $F$ Theory}

If one is to maintain space-time supersymmetry, the maximum number of 
space-time dimensions which is allowed for a consistent field-theory 
interpetation is $d=11$, if one accepts that there 
should be no massless particles with spins higher than two~\cite{nahm}.
However, there is another stringy reason which restricts the 
maximum number of dimensions: $d \le 11$.
As we discussed recently in~\cite{emnmonopdbr}
and reviewed briefly in section 1,
a world-sheet theory with monopoles leads to $D$
branes in target space. This was seen by noting that
non-conformal monopole deformations in the region $9 < d <11$ 
induce cuts in the correlation functions (\ref{twopoint}) of the 
defect vertex operators with the vertex operators describing 
the excitation of target-space matter fields, leading to 
an effectively open world sheet. Dirichlet boundary conditions
can then be placed along the cuts 
by canonical $T$-duality
transformations in the appropriate $\sigma$-model path 
integral~\cite{emnmonopdbr,otto}, to obtain solitonic 
$D$ branes in target space.  

It was apparent from our non-critical string construction 
that the target space of such a $d$-brane 
is {\it a priori} $11+1=12$-dimensional. This is the maximum number of 
dimensions in which supersymmetry can exist, {\it if
the 12-dimensional space time has } $(10, 2)$ {\it
signature}~\cite{duff,bk}. This is because such a signature
admits a spinor which is both Majorana and Weyl~\cite{duff,nishino}.
Due to the Liouville dynamics~\cite{aben} for $d > 9$, this implies that 
the conformal point d=11 must have (10, 1) signature for supersymmetry 
to appear in the full non-critical string, viewed as a critical one 
in a different target-space dimension. We also note that, among
Minkowski space times admitting $p$ branes, the 11-dimensional 
manifold  is the {\it maximal} one admitting super $p$ branes~\cite{duff}.

Our construction of $D$ branes, via topological defects
on the world sheet of the non-critical Liouville string, 
provides a natural framework
for the appearance of such a 12-dimensional 
theory, for which the 11-dimensional supergravity appears as a fixed 
conformal point. We therefore conjecture that the above 
construction may lead to
a $\sigma$-model description of $F$~theory. 

In this construction, the critical 
string theories appear as conformal fixed points - effectively in
`equilibrium' -
of the Liouville string -
non-critical and `non-equilibrium' in general~\cite{emnmonopdbr}) -
which describes
the bulk of theory space. 
Each fixed point in the space of string theories
has its own time-like coordinate $X^0$, which is a {\it reversible}
coordinate in the Einstein sense, so that
a fixed-point theory is Lorentz invariant.
In addition, in the bulk of the
theory space there is a second time-like coordinate, 
the Liouville field $\phi$, with respect to which evolution is
{\it irreversible} in general~\cite{emn}. This means that
the general $d+1$-dimensional target-space $F$ theory is
`non-equilibrium', and does
not in general have a simple field-theoretical
interpretation~\footnote{This is consistent with the fact~\cite{nishino}
that the 12-dimensional supergravity model is not Lorentz invariant.}.
This type of deviation from conventional Lorentz-invariant
field theory cannot be ignored in general, since
defect deformations cause the theory to fluctuate
away from the conformal points, without necessarily driving the theory
to a new fixed point. Our proposal for managing
the appearance of these two times is to identify the
zero modes of the {\it a priori} distinct quantum fields
$X^0$ and $\phi$~\cite{emn} in the neighbourhood of each fixed point. We
have demonstrated that
this identification reproduces correctly the metric of
the $1+1$-dimensional string black hole~\cite{emn}, and implements
correctly energy-momentum conservation in closed-string
scattering off a $D$-brane target, including quantum recoil
effects~\cite{diffusion}.
We believe that this approach may cast some light on the mysterious
structure of $F$ theory.

%\newpage
\vspace{0.1in}

{\bf Acknowledgements}

The work of D.V.N. is supported
in part by D.O.E. Grant
DEFG05-91-GR-40633.

\end{document}